# Forecasting Change in Conflict Fatalities with Dynamic Elastic Net

Fulvio Attinà[1] (fulvio.attina@unict.it)

Marcello Carammia[1] (marcello.carammia@unict.it)

Stefano M. Iacus[2] (siacus@iq.harvard.edu)

[1] Department of Political and Social Sciences, University of Catania

[2] Institute for Quantitative Social Science, Harvard University

Corresponding author: Marcello Carammia

## Abstract

This article illustrates an approach to forecasting change in conflict fatalities designed to address the complexity of the drivers and processes of armed conflicts. The design of this approach is based on two main choices. First, to account for the specificity of conflict drivers and processes over time and space, we model conflicts in each individual country separately. Second, we draw on an adaptive model – Dynamic Elastic Net, DynENet – which is able to efficiently select relevant predictors among a large set of covariates. We include over 700 variables in our models, adding event data on top of the data features provided by the convenors of the forecasting competition. We show that our approach is suitable and computationally efficient enough to address the complexity of conflict dynamics. Moreover, the adaptive nature of our model brings a significant added value. Because for each country our model only selects the variables that are relevant to predict conflict intensity, the retained predictors can be analyzed to describe the dynamic configuration of conflict drivers both across countries and within countries over time. Countries can then be clustered to observe the emergence of broader patterns related to correlates of conflict. In this sense, our approach produces interpretable forecasts, addressing one key limitation of contemporary approaches to forecasting.

**Acknowledgements**: We are grateful to Francesco Bonaccorso for excellent assistance in data processing. We are thankful to the ViEWS team, and especially Paola Vesco, for support and advice. We thank two anonymous reviewers and the journal editors for their comments that greatly improved the manuscript.

# 1. Introduction

This article presents an approach to forecasting change in conflict severity based on a Dynamic Elastic Net (DynENet) model (Carammia et al., 2022). The approach was designed to respond to the call of the ViEWS project (Hegre et al., 2022). Our study joins the empirical tradition of the quantitative study of internal and international violent conflict and war. In the early time of the quantitative analysis of violent conflict, from the 1960s to the 1990s, new techniques were invented, such as the 'international event analysis', and archives of data were generated mostly for the empirical testing of international theories such the COW, WEIS, and DON databases. Social constructivism and critical studies have restricted the growth of such research and diverted efforts towards the interpretivist study of political process and phenomena. However, quantitative analysis has never ceased to be one of the useful tools for conflict research and, undeniably, also a powerful tool for policy science and research applied to policy decisions. This certainly applies when decision-makers face an internal and international conflict, whether peaceful or violent, and need practical knowledge to predict its future evolution.

Conflicts are complex processes. Actors have a high degree of discretion as common authority is delegitimized, but in general drivers are numerous (Buhaug et al., 2014) and diverse enough to include political factors Hegre (2014), ideology (Sanín & Wood, 2014), foreign interventions (Kathman & Wood, 2011), the climate (Koubi, 2019; Nordås & Gleditsch, 2007) and ethnicity (Denny & Walter, 2014), in addition to territorial factors (Toft, 2014). Therefore, drivers vary largely over space (across countries) and time (across and within countries), forming complex context-dependent driver configurations. The size of their effects varies wildly, and so do interactions and dependencies (D'Orazio, 2020), meaning that they can have largely different impacts across contexts. Policymakers take into account various factors when responding to conflict-driven events: responses are not necessarily tit-for-tat reactions generating escalation processes. After conflicts start, the level of violence and the severity of the process are influenced but are not dependent on the configuration of the circumstances in which the conflict is generated.

In sum, the conflict severity is but one of the dimensions the conflict takes on under the effect of the complexity of the international and domestic political systems. The complexity of the international system makes conflict forecasting a challenging exercise, one that is the object of a growing scholarship (see e.g. Brandt et al., 2011; Colaresi & Mahmood, 2017; Schrodt, 1991; Ward et al., 2010; Weidmann & Ward, 2010; for a review, see D'Orazio, 2020; Hegre et al., 2017; Schneider et al., 2010). Our forecasting approach builds on these efforts.

We take complexity seriously and design our approach accordingly, bringing some novel contributions to the field of conflict forecasting. First, to address the specificity of conflict drivers and processes across countries, we model conflicts in each country separately. Contrary to standard approaches, we do not assume that a single model can predict conflict in any context. Second, we base our approach on an adaptive model – DynENet – that is able to efficiently select relevant variables among a very large set of potential predictors. For each individual country, our model selects only the variables relevant to that specific context and drops all the others. This is done in a dynamic fashion, as potentially different sets of drivers are selected to forecast different steps (months) ahead in our time series.

In addition to being consistent with the complexity of conflict processes, the adaptive nature of our modelling approach brings another added value. Because for each country our model only selects the variables relevant to predict conflict severity in the country analyzed, the retained predictors can be observed to analyze the constellations of conflict drivers across countries, and within countries over time. Countries can then be clustered based on their prevalent conflict drivers, to investigate the emergence and characteristics of higher-level patterns. Therefore, our approach delivers interpretable forecasts, although primarily data-driven ones, and below we provide some brief examples of how this is achieved. In this way, our approach addresses one significant limitation of forecasting approaches, in which accuracy is often achieved at the expenses of interpretability (Hegre et al. 2022).

The next section illustrates the data and the design of our approach. The analysis first presents a summary overview of the forecasts generated for the competition, focused on the variation in conflict fatalities in the six months between October 2020 and March 2021, two to seven steps ahead of the data that we were provided for this exercise. It then evaluates the performance of our model in the test set. The efficiency of DynENet is compared against a simpler LASSO model, and the contribution of different data sources is analyzed. The analysis also briefly discusses the performance of DynENet compared to the set of models that took part to the competition, as well as a benchmark Random Forest model. Finally, the analysis provides summary interpretations of the forecasts for a selection of countries – Cameroon, Nigeria, Rwanda and Senegal – and a cluster analysis of countries based on their conflict drivers. The conclusions reappraise the results of this exercise and discuss some strengths and limitations of our approach, also in the light of possible future developments.

## 2. Data

Our approach draws on the data provided by the ViEWS project, to which it adds an additional set of event data.

### *1. ViEWS data*

Our first source of data was the ViEWS project. In preparation for the forecasting competition, ViEWS released data at https://views.pcr.uu.se/download/datasets/ in three waves, on 24 March, 27 April, and 29 September 2020. At each wave, the dataset included different variables (respectively circa 400, 650, and > 2000). We worked with the most recent release, which was also the largest by far. However, because the data were released just a few days before the deadline, which meant we did not have time to carefully inspect the data included there, we only selected the 653 variables which were included in the April 2020 release. This data release aggregated indicators provided by a number of sources, including:

- 48 GED variables taken from the UCDP Georeferenced Event Dataset, which record events of lethal organized violence occurring at a given time and place (Pettersson & Öberg, 2020; Sundberg & Melander, 2013).
- 8 ACLED variables from the Armed Conflict Location and Event database (Raleigh et al., 2010).
- 363 World Bank's World Development Indicators (WDI), which include data on a range of such structural indicators as agriculture and the economy, education, health, infrastructure, poverty, public and private sectors, social development and trade, among others (World Bank, n.d.).
- 129 political regime variables from the Varieties of Democracy (V-Dem) project (Coppedge et al., 2011).
- 51 FVP variables, a stacked dataset compiled by the ViEWS for other projects and including variables on ethnic relations, shared socio-economic pathways, and democratic governance Hegre et al. (2019).
- 49 variables from the Rulers, Elections, and Irregular Governance (REIGN) dataset, including data on the characteristics of world leaders, political institutions and political regimes, election outcomes and announcements, and other irregular events as coups, coup attempts and other violent conflicts (Bell, 2016).
- 5 ICGCW variables from the conflict tracker of the International Crisis Group (https://www.crisisgroup.org/crisiswatch).

for a total of 653 covariates taken from the ViEWS dataset.

Overall, the data covered the period from January 1990 to August 2020, at the monthly level.

*2. GDELT event data*

On top of the ViEWS data, we added a source of event data. For this, we relied on the Global Database of Events, Language, and Tone (GDELT) project (Leetaru & Schrodt, 2013), a repository of 316 types of geolocated events reported in the world's broadcast, print and web media, in 100 languages.[1] We used GDELT 1.0 data, which are updated on a daily basis; but we aggregated them on a monthly level for the purposes of this particular analysis. Note that GDELT 2.0 data are updated every 15 minutes and could potentially feed a near real-time system.

---

[1] See Web Page at: https://www.gdeltproject.org. GDELT "monitors the world's broadcast, print, and web news from different countries in over 100 languages and identifies the people, locations, organizations, counts, themes, sources, emotions, counts, quotes, images and events driving society daily".

Nuanced event data such as GDELT have the potential to contribute both to the accuracy and, in case of interpretable forecasts such as ours, to the understanding of conflicts. There are known potential reliability issues associated to machine-coded event data. In future developments of this project we plan to test other sources such as UTD event data (Kim et al., 2019; Solaimani et al., 2016).

We extracted from the GDELT database only those events that occur in the lead paragraph of a document (coded as 1 on the "IsRootEvent" variable in GDELT). Single events can be covered multiple times by different media outlets and therefore occur multiple times in the GDELT data, so we extracted individual events from the overall media coverage and include each event only once in our data.

For the event categories, GDELT relies on the Cameo scheme[2] (Gerner et al., 2002). Cameo is a hierarchical codebook, meaning that events can be aggregated up to the level of macro-categories. We built our indicators using the 20 Cameo event macro-categories.[3]

Individual events differ in the extent to which they may generate conflict. The GDELT codebook[4] has two variables to account for this. One is the Goldstein scale, which goes from -10 to +10, and is meant to capture precisely the conflict-generation potential of each event (Goldstein, 1992). The other one is QuadClass, which is used to distinguish between different degrees of 'action' in each event and has four values.

For each event macro-category, we then developed indexes identified by the following initials:

---

[2] CAMEO scheme http://data.gdeltproject.org/documentation/CAMEO.Manual.1.1b3.pdf.

[3] Namely: statement; appeal; express cooperation; consult; diplomatic cooperation; material cooperation; aid; yield; investigate; demand; disapprove; reject; threaten; protest; show force; reduce relations; coerce; assault; fight; use unconventional violence.

[4] See GDELT codebook http://data.gdeltproject.org/documentation/GDELT-Event_Codebook-V2.0.pdf.

- `m` = number of mentions (in the media). this is the natural GDELT value assigned to events under variable NumMentions
- `e` = event. Here we filter out media coverage to represent the 'pure event'. Because the single observation in the GDELT dataset is a single event (identified by the variable GlobalEventID), we simply reduce NumMentions=1 for each individual observation
- `eg = e*g`. Event weighted by Goldstein scale (-10, +10).
- `eq = e*q`. Event weighted by QuadClass scale (1=Verbal Cooperation, 2=Material Cooperation, 3=Verbal Conflict, 4=Material Conflict).[5]
- `egq = e*g*q`. Event weighted by Goldstein e QuadClass.

Our covariates therefore include 100 series generated from the GDELT database: 20 event macrocategories * 5 different weights.

## 3. Methods

### 1. Forecasting model

The outcome to predict is in the change in the number of fatalities per month from state-based conflict. More precisely, participants to the forecasting competition were required to predict the step specific delta of the `ln_ged_best_sb` variable from UCDP-GED data (Sundberg & Melander, 2013), which is the change over *s* [2, 7] months in the natural log of 1 + the UCDP GED best estimate of deaths from state-based conflict events in a country-month (Hegre et al. 2022, Vesco et al. 2022).

We estimated a separate Dynamic Elastic Net Model for each individual country. For the purpose of this competition, we did not test different training and calibration periods. Therefore, for this first implementation of our approach we simply relied on the periodization suggested by the ViEWS project (Hegre et al. 2022). We set $y_{s[2,7]}$ as the outcome variable, and used all the months available in the past as a training set. We plan to test more alternative specifications of training and calibration periods in the future.

---

[5] To produce the forecasts presented in this paper, we used these values for the QuadClass weight: 1=Verbal Cooperation, 2=Material Cooperation, 3=Verbal Conflict, 4=Material Conflict. One anonymous reviewer pointed out that this does not accurately follow an ideal ordering from less to more 'conflictual'. This may be a subtle point, but it is certainly a correct one. Probably a better ordering could be obtained by trading weights 1 and 2, so that the scale would go from least (material cooperation) to most conflictual (material conflict). Or, we could have just used QuadClass 3 and 4 (conflict related) as possible weights. Because we designed the forecasting approach described in this article as part of the forecasting competition, we cannot revise ex post any potential flaws of the approach, which we plan to address in future implementations. Even though not an entirely accurate measurement choice, however, within the framework of a data-driven forecasting model this does not represent a fundamental bias, in our view (as it would instead be in the case of an explanatory model). An inaccuracy or error in the construction of a single indicator will reflect in the performance of the model, but not necessarily affect the reliability of the overall approach. Forecasting models are primarily evaluated by their performance (Toshkov 2016, 33). More specifically for what concerns our particular approach, as we mentioned above and further explain below, our algorithm only selects single (weighed) variables if they have some predictive relevance to the particular forecast that is being produced. In practice, this means that if one particular (weighed) variable has no predictive relevance (or it correlates to a similar variable with stronger predictive strength) it is dropped from the model (in favour of the variable with stronger predictive relevance). If GDELT variables weighed byQuadClass are selected in the model, that indicates that they are effective predictors of conflict in the particular context where the model retains them (although perhaps not as effective as they could have been, had they been more accurately designed).

Our procedure uses the training set to estimate a DynENet model, as well as for further cross validation to minimise the forecast error along the training period. DynENet (Carammia et al., 2022) is the Elastic Net method (Zou & Hastie, 2005) ran on a dynamic window. Being an adaptive method that mixes LASSO-type (Tibshirani, 1996) and Ridge-type (Hoerl & Kennard, 1970) estimation, the DynENet can take into account hundreds of variables for each country. It has the advantage of finding the most parsimonious model for each country; at the same time, the model also takes into account collinear variables as in Ridge regression.

DynENet is a relatively new type of regularization method which tries to perform model estimation and model selection in just one run. Suppose we want to estimate a linear model of the form $y = X\beta + \varepsilon$, where $X$ is the matrix of regressors and $y$ is the dependent variable of interest (in this case, change in the number of fatalities). In the context of this project, we have a huge number of regressors and relatively few observations, which prevents us from estimating a new model. Regularization methods, like DynENet, are also meant for dimensionality reduction, i.e., they estimate some of the beta coefficients as zero:

$$\min_{\beta_0,\beta} \left\{ \frac{1}{N} \sum_{i=1}^{N} w_i l(y_i, \beta_0 + \beta^T x_i) + \frac{\lambda}{2} \left[ (1-\alpha)||\beta||_2^2 + 2\alpha ||\beta||_1 \right] \right\}$$

where, $w_i$ are weights for observations ($w_i$=1 by default), $l(\cdot)$ is a loss function, normally the classical least squares contrast function, $\lambda$ is a penalty factor and $\alpha$ is a tuning parameter.

More precisely, in our case the forecast function in DynENet takes this form:

$$\min_{\beta_0,\beta}\left\{\frac{1}{N-k}\sum_{i=1}^{N-k}(y_{i+k}-\beta_0-\beta^T x_i)^2 + \frac{\lambda}{2}\left[(1-\alpha)||\beta||_2^2 + 2\alpha||\beta||_1\right]\right\}, for\ k=2,\ldots,7$$

The DynENet model thus has two tuning parameters: $\alpha$ and $\lambda$. The first one, $\alpha$, is set to 0.5 in our approach, which means LASSO (L1-penalty) and Ridge (L2-penalty) estimations are equally weighted in the loss function of the optimization problem. The parameter $\lambda$, the adaptive scaling factor for the penalties, is first estimated using cross-validation over a sequence of possible values that minimise the forecast MSE in the training data and, at the same time, guarantee that the explained deviance of the model is at least a given value (e.g., 50% or 75%). The resulting 'optimal forecasting value' is then used in the DynENet penalty function.

The tuning parameter $\lambda$ is quite important. The larger this number, the stronger DynENet will shrink the estimated coefficients to zero, in a potentially artificial way. To take into account this potential source of bias, the dynamic net considers a grid of different values for $\lambda$ and estimates the penalized regression model for a particular choice of $\alpha$ (in our case 0.5). Then $\lambda$ is automatically selected by cross-validation. For $\lambda = 0$, the formula becomes the usual OLS (least squares estimation) approach. For $\lambda = 1$, this method becomes the so-called LASSO regression model, i.e., when trying to minimize the squared residuals from the model, the L1-penalty ( $||\beta||_1$ = sum of the absolute values of the regression coefficients) is added forcing some of the coefficients to be estimated as zero.

For $\alpha = 0$, this model becomes the Ridge regression model, *i.e.,* the classical regression with shrinkage for robust error estimation. For $\alpha = 0.5$ the model is simply called Dynamic Elastic Net (DynENet). Using both L1 and L2 penalty at the same time is a good compromise in terms of prediction. In fact, LASSO regression tends to keep only one among highly correlated subsets of regressors, discarding all the others. With $\alpha = 0.5$, the result of the regularization also takes into account the correlation among the regressors, which results in a sort of "mean" effect of all variables that matter even though correlated among them. Note that DynENet is also a variance shrinking method, which implies that the standard errors of the coefficients of the selected variables are relatively small compared to, e.g., linear regression.

In practice, for each forecasting step ahead of the outcome variable $y_{s[2,7]}$ the system proceeds as follows:

- Set up an Elastic Net model;
- Perform model estimation using the data for all points (months) available in the past;
- Apply the estimated model to predict the outcome variable;
- Repeat the procedure for each *s[2, 7]* step ahead.

Because for each country/step ahead DyNEnet retains a limited and potentially different set of predictors, and it discards the other variables, we can observe the retained predictors to trace how covariates of conflict fatalities vary both across countries and within countries over time. We implement this driver analysis by running *ex post* Random Forest models to compute variable importance (Hastie et al. 2009, 593-595). In practice, for each country, we take the variables selected by DynENet and – using only those variables – we compute variable importance measures to rank the predictors based on their relative contribution to each country/month forecast. We then visualise the ranked variables in separate country heatmaps, where we also map their changing relevance over time.

In 9 out of 35 cases, when the Elastic Net algorithm did not converge due to too scarce variability in the outcome variable or in the covariates, we replaced the prediction of the outcome variable with its historical running mean. This is a fallback strategy that we consider part of the DynENet method. For this reason, we have 26 countries in which predictors are dynamically selected.

In a further step, to investigate the existence of higher-level patterns, we use a k-means algorithm to cluster countries based on the similarities and differences among country-specific conflict predictors for the 26 countries for which we have dynamically selected predictors. We include sample forecast interpretations for some countries, as well as cluster analyses, in the discussion of the results that follows, which starts with a presentation of the results of the forecasts.

## 4. Analysis

DynENet generated forecasts for 35 out of 52 African countries. In most cases, when forecasts were not generated that was due to the large number of zeros or missing observations in the training set. Because the model is fit separately for each country, DynENet is not able to generate a country-specific prediction if a country has too limited or no past data on the outcome variable or covariates. A possible approach to overcome this limitation of country-specific model estimation would be a two-step procedure: in the first step, match the covariates of a target country to those of a country for which a model has been successfully estimated, using e.g, a coarse algorithm like CEM (Iacus et al., 2012, 2019); then, in the second step, apply the selected model to the covariates of the target country to predict conflict fatalities. This extension is in the pipeline for future developments of the proposed method.

Predicted changes in conflict fatalities at steps 2 (October 2020) and 7 (March 2021) are shown in Figure 1. The countries with the highest expected increase in conflict fatalities are Niger, Libya, Chad, Algeria, Egypt and Burkina Faso. For these countries, the total expected delta in the target variable over the forecasting period ranges between 13.9 and 3.6. Sudan, Cameroon, Senegal and Angola (with a six-month expected delta comprised between one and two) complete the top-10 countries with the highest expected increase in conflict severity.

[ FIGURE 1 ABOUT HERE]

Sixteen countries are forecasted to be rather stable. Among them, Uganda, Tanzania, Congo and Cote D'Ivoire have an expected delta between 0.9 and 0.5. Rwanda, Mauritania, Kenia and Djibouti have an expected delta between 0.35 and 0.1. Virtually no change is expected to take place in Eritrea, Zambia, Gambia, Benin, Malawi, Namibia, Ghana and Lesotho. Conflict fatalities are instead expected to decrease in the Democratic Republic of Congo, Somali, Mali, South Sudan, Nigeria, Burundi, Ethiopia, Central African Republic, and Mozambique, with values ranging between -1.3 in Congo DR and -20.8 in Mozambique.

## 1. *Evaluation of the forecasting performance*

Here we present an evaluation of the performance of our forecasting models. We present the main error statistics employed in the competition, discuss the overall performance of DynENet compared to the models that took part to the ViEWS competition, and then provide some in-depth analyses of DynENet's performance in the test set. The first metric selected for the ViEWS forecasting competition is the Mean Square Error (MSE), a standard metric for forecast evaluation. The MSE is calculated as:

$$\text{MSE}_{(y,\hat{y})} = \frac{1}{n}\sum_{t=1}^{n}(y - \hat{y})_t^2$$

where $\hat{y}$ is the expected value and $y$ is the true value.

Alongside of the MSE, Vesco et al. (2022) also proposed the Targeted Absolute Distance with Direction Augmentation (TADDA):

$$\text{TADDA}^{(M)} = \frac{\sum_{i=1}^{N}|y_{\Delta,i} - f_{\Delta,i}| + |f_{\Delta,i}|\,\text{I}\left[y_{\Delta,i}^{\pm} \neq f_{\Delta,i}^{\pm}\right]\text{I}\left[|y_{\Delta,i} - f_{\Delta,i}| > \epsilon\right]}{N}$$

We estimated two versions of TADDA with different specifications of ϵ. In TADDA–A, ϵ for step k is sd(T$_k$), where the "true difference at step k", $T_k = y(t+k) - y(t)$. In TADDA–B, ϵ for step k is $z_\alpha * sd((T\_k))/\sqrt{n}$, with α = 0.995 and z$_α$ the alpha quantile of the standard Gaussian distribution.

DynENet performed overall well compared to the other models that took part to the forecasting competition, although it did relatively better in the test set than with the true future partition. A comprehensive analysis is provided in Vesco et al. (2022), where all (nine) models are compared against each other, as well as against a benchmark Random Forest model, a no-change scenario, and an ensemble of all models – for a total of twelve compared models. In the test set, DynENet performed quite well (fourth) in terms of MSE, and it largely outperformed the benchmark Random Forest model. Among all models, the predicted variance of DynENet was the second closest to the variance of actual values, and it had among the best scores for forecasts of negative changes. Forecasting approaches were also compared in terms of Model Ablation Loss (MAL) at steps 2, 4, 7, obtained by observing how the performance of the ensemble of the predictions changes when dropping a specific model. DynENet was the model generating the highest ablation loss at steps 2 and 4, and the second highest at step 7. Clearly, DynENet's unique contribution to the 'wisdom of the crowd' of the ensemble of models was highly important among forecasts of the training set.

DynENet was relatively less efficient forecasting the true future. In terms of MSE, the most established performance metrics, it went down from the fourth to the seventh best-performing model (although with forecasts of the true future, the MSE of the eight higher-ranking models were rather close to each other, compared to the remaining ones). Still, DynENet's performance was well above the median performance of competing models on twelve out of thirteen evaluation metrics; much closer to the best-performing model than the median in most metrics; and first or second in terms of (standard deviation or mean) forecast calibration (see Figure 6 and the related discussion in Vesco et al. 2022).

We carried out a more in-depth analysis of DynENet's performance in the test set (January 2017 – December 2019). We use MSE and TADDA to a) evaluate the performance of our forecasting model; b) test it against a LASSO model, that we take as a relatively simpler benchmark; and c) evaluate the loss resulting from the ablation of GDELT data from our model. Figure 2 shows that, on the average (that is, when evaluation metrics are averaged across countries, for each of the months included in the test set), DynENet outperforms LASSO at all steps. However, DynENet seems to perform slightly better when GDELT data are *not* included in the model. These results are consistent across all evaluation metrics: MSE, TADDA-A, TADDA-B.

**[FIGURE 2 ABOUT HERE]**

Table 1 quantifies the efficiency of DynENet over LASSO, measured as the ratio between error statistics (MSE or TADDA A or B) of the two models. A value of 1 would indicate that the two models have the same performance. Lower values indicate that DynENet outperforms LASSO, while higher values would indicate the opposite. On the overage (across countries), DynENet outperforms LASSO at all steps into the future. It does so by a relatively small margin; although the efficiency ratio tends to decrease over forecasting steps, meaning that the margin of efficiency of DynENet over LASSO increases along forecasting steps. Apart from the small loss of efficiency, DynENet is qualitatively better than LASSO for the purpose of interpretable forecasts, like ours. If there are two or more correlated variables, LASSO just selects one. But if although correlated those variables are all important, DynENet thanks to the Ridge component will retain them all, in practice averaging the coefficients. This can significantly improve the interpretation of the model. Take as an example the purchase or provision of weapons, which clearly may correlate with conflict-related covariates. LASSO would then choose either one or the other, while DynENet would usually weigh them efficiently, thus allowing for higher interpretability of the results.

**[TABLE 1 ABOUT HERE]**

Another distinctive element of our approach lies in the addition of GDELT event data to the set of 653 variables included in the ViEWS dataset. As a measure of the contribution of including GDELT data over and beyond those variables, Table 2 shows the Data Ablation Loss (DAL) which results from removing GDELT data from the model. Similar to the Model Ablation Loss in Vesco et al. (2022), GDELT DAL is measured as:

$$DAL_{GDELT}^{EM} = EM^{full} - EM_{GDELT}^{abl}$$

for each evaluation metric (EM), that is, for MSE, TADDA–A, and TADDA–B.

**[TABLE 2 ABOUT HERE]**

Table 3 shows the average MSE at forecasting step 7, along with the DynENet/Lasso efficiency ratio and the GDELT DAL, for single countries. We show figures for forecasts at step 7 to provide a conservative test. While errors do tend to increase along the forecasting horizon, for each country their magnitude is largely consistent and comparable among steps.

**[TABLE 3 ABOUT HERE]**

The full DynENet model is equally or more efficient (efficiency rate ≤ 1) than Lasso for most countries. Lasso is slightly more efficient in forecasting Algeria, Lybia, Mali and Nigeria, although the efficiency improvement is even smaller if GDELT data are removed from DynENet. In turn, the full DynENet advantage is largest for the Democratic Republic of Congo, Cameroon, Egypt. Removing GDELT gives DynENet an additional advantage in forecasting change in conflict fatalities for Kenya and Mali, and also makes the forecasts for Nigeria more (rather than less) efficient than LASSO (see Figure 3).

[FIGURE 3 ABOUT HERE]

Figure 4 shows for each country the ablation losses resulting from removing GDELT data from the full model. DAL is negative (meaning that there is a performance loss from removing GDELT data) only for Burkina Faso, Congo DR, and Niger. Several other country forecasts improve by a tiny margin from removing GDELT from the model, whereas the forecasts for Libya, Mali, Mozambique and Chad improve more markedly.

[FIGURE 4 ABOUT HERE]

In sum, the ablation of GDELT data results in a small (yet consistent across metrics and forecasting steps) overall gain in the performance for most countries; it produces a loss only for a minority of countries. This requires further investigation. It may indicate the need to better calibrate the weights that we applied to GDELT events, and/or to only include those weights that optimise the forecasting performance (see section on GDELT data above, and footnote 5). Future implementations of the model could also test event data sources alternative to GDELT.[6] However, a (marginal) performance loss may still be acceptable in exchange for the information that is obtained from this data feature, which can be used for interpreting the forecasts – as we explain in the next section.

*2. Forecast interpretation. Case studies*

The adaptive nature of our model, combined with the by-country fitting approach, means that our approach has the added value of producing *interpretable forecasts*. In practice, for each country the retained predictors provide an individualised description of drivers of conflict, and of their change over time. Here we present as an example a short analysis focused on four countries: Cameroon, Nigeria, Rwanda and Senegal. The countries selected provide a diverse combination regarding the magnitude and direction of forecasted change in conflict fatalities; the forecasting performance of our model in the test set; and the number and class of predictors retained by the model (see Table 4 and Figure 5).

**[TABLE 4 ABOUT HERE]**

**[FIGURE 5 ABOUT HERE]**

In the discussion that follows, driver analyses for the selected countries are visualised in separate heatmaps (Figures 6-9). In each figure, the horizontal axis is a timeline of the months included in the training set, whereas the vertical axis includes the set of relevant predictors selected by DynENet for the country analyzed. Each cell in the map includes a measure of the importance of each predictor at each month, represented through colors and ranging from from zero (white – variable not included) to 1 (red - variable most important). The heatmaps denote some recurring features of the analyzed processes. The effect of single variables tends to persist for several months, which indicates some temporal structure that the model is able to detect. However, the effect of single variables or clusters of variables also tends to vary over time and in some cases to disappear entirely, which denotes change in the drivers of conflict – again, change that the model is able to recognise.

---

[6] Alternative sources of event data exist and in academic research, compared to other public or private sectors, they are more frequently employed than GDELT. We selected GDELT for our model because we wanted to design a forecasting system employing (at least some) data that would be updated in near real time, so as to feed a potentially live forecasting system also including real time alert functionalities. Other options would include the UTD Event Data, which were recently broadened to include near real time data (Kim et al., 2019; Solaimani et al., 2016).

*Cameroon*

Our model forecasted a total delta of 1.931 in the target variable for Cameroon between October 2020 and March 2021. This is the eight largest forecasted variation in the set of countries, with an across-country average of -0.036. The MSE for Cameroon was relatively high in the test set, where the DynENet efficiency ratio over LASSO was marked at 0.87. The model retained a high number of predictors from several data classes (Table 4 and Figure 6). The random forest model ranked GED highest among predictors, but a significant number of WDI structural indicators also were included. Indeed, a subset of three GED and two WDI variables were core predictors for most months between October 2014 and December 2019. This points to a forecast driven primarily by structural factors, as well as some inertia in conflict fatalities captured by GED data and by some ACLED variables.

**[FIGURE 6 ABOUT HERE]**

Some events data also were selected among predictors, as indicated by such GDELT variables as '`showforce`', '`demand`', or '`threaten`', which were retained by the model in 2015 and the beginning of 2016. GDELT variable 'assault' also was included among predictors for most of the training period, although not highly ranking. GDELT DAL for Cameroon is positive, meaning that in the test set the performance improved slightly when GDELT data were excluded. As one of the cases whose forecast improves without GDELT data, but also one where GDELT enriches the interpretation of forecasts, Cameroon provides a good example of the potential trade-off between forecasting accuracy and analysis of conflict determinants. Considering the relatively minor precision loss, we see value in including event data in the model.

*Nigeria*

Our model predicts a marked downward trend in the number of conflict fatalities in Nigeria, with a delta of -3.907 over the forecasting period – the fifth largest expected decrease. A larger decrease was only forecasted for Burundi, Ethiopia, Central African Republic and Mozambique. However, if we look at levels instead of change, Nigeria is still forecasted as the country with the second highest level of conflict fatalities – despite the negative trend. Only Somalia ranks higher. In terms of forecasting performance, the MSE for forecasts of Nigeria in the test set was lower than Cameroon, although still significant at 0.915.

DynENet retained a large number of predictors (Figure 7). Among the selected variables, WDI indicators were by far more frequent and higher-ranking. The second most represented data class was V-Dem, which was also consistently present throughout the training period. Together, WDI and V-Dem sum up 43 out of 48 variables. Some GED variables were either consistently retained but not high ranking, or highly relevant but only included for few months in the first half of the series. ACLED and FVP variables only had a marginal presence. This indicates that in the case of Nigeria conflict fatalities were primarily driven by structural variables, either socio-economic and demographic (WDI) or political (V-Dem) ones.

**[FIGURE 7 ABOUT HERE]**

Even in the case of Nigeria, DynENet was more efficient than LASSO (0.816). Interestingly, the large number of predictors, and their intermittent relevance over time (as indicated by 'patchy' nature of the chart), did not result in poor forecasting performance in the test set.

*Rwanda*

Our model predicts an almost flat tendency for Rwanda, with a total six-month delta of 0.36. Not only is change in conflict fatalities forecasted as very low; the forecasted levels of fatalities in Rwanda also are among the lowest in Africa. Perhaps unsurprisingly, the model retained a very small number of predictors, selected over a small number of observations as well (Figure 8). Indeed, at each point in time one variable alone was sufficient to generate the forecast: one GDELT variable related to contestation events between 2014 and mid-2018; and then mostly the GED variable, except for two occurrences of V-Dem variables. While this would require proper analysis, the model seems to describe a context where primarily political factors drove conflict, along with some inertia in conflict fatalities.

**[FIGURE 8 ABOUT HERE]**

The limited number of retained variables was sufficient to generate highly reliable forecasts in the test set, with an MSE of 0.357. The efficiency ratio indicates no efficiency gain over Lasso. Apparently, in the case of Rwanda the DynENet set the λ parameter to 1, and the model worked as a LASSO model forcing many coefficients to zero. The back test indicates that GDELT data were included at no cost for the reliability of forecasts (DAL = 0). Arguably, however, the inclusion of event data increased the interpretability of the results.

Senegal

DynENet forecasted a moderate variation in Senegal, with a six-month predicted delta of 1.09 making it the country with the ninth-largest expected increase in Africa. The number of predictors retained was relatively limited; but it was sufficient to generate highly reliable forecasts in the test set, with an MSE of 0.051 placing Senegal among the top-10 more reliable forecasts. As was the case for Rwanda, there was no efficiency advantage of DynENet. Together with the limited number of predictors retained, this indicates that DynENet effectively operated as a LASSO model.

Four WDI and two GDELT variables were retained (Figure 9). Only the GDELT variable 'threaten' was selected in 2014 and 2015, both as a raw event and weighed by the Goldstein index. Since 2016, the model only selected WDI indicators – mainly poverty indicators such as maternal mortality (`sh_sta_mmert_ne`), stunting (`sh_sta_stnt_ma_zs`), and malnutrition (`sh_sta_stnt_ma_zs`). Even in the case of Senegal, GDELT data did not affect the reliability of forecasts; but they added to the understand of its drivers.

**[FIGURE 9 ABOUT HERE]**

*3. Forecast interpretation. Country clusters*

By-country fitting is central to the design of our approach. However, as a result of international-level factors as well as broad regularities in conflict dynamics, we should expect some degree of structure in conflict processes across countries. Cross-country patterns are not detected by single-country models, but they can be examined by clustering countries based on the similarities and differences among their specific conflict predictors. Note that this cluster analysis is only performed on the 26 countries where DyNEnet dynamically selected predictors (see design section above).

We proceeded as follows. We first developed summary scores of overall predictive importance of single variables for each country, along the entire period observed. We measured the overall predictive importance of a variable for a country as the product of the number of times that variable was selected by DynENet as a relevant predictor for that country, and its average rank as predictor based on the Random Forest variable importance measure. Having obtained for each country-specific predictor the associated overall importance score, we had to aggregate predictors. Possible aggregation criteria would include source (e.g. GDELT, ACLED, V-Dem, etc.), content (e.g. violent events, institutional factors, structural factors, etc.), or both (e.g. events/ GDELT, events/ ACLED, structural factors/ FVP, structural factors/ WDI, etc.). Here we show results for the latter aggregation criterion, *i.e.* by source and content; but we tested other aggregation criteria as well.

Having aggregated variables, we should then decide how to aggregate the previously created scores of overall predictive importance. We tested two approaches, one based on the average importance score for each variable, the other one – which we report here – based on a binary measure (1 if the variable was a predictor for the country, 0 if it was not). Finally, having measured the overall predictive importance of each variable/country, aggregated predictors onto codes (source/content), and assigned to each code for each country a value of one if it was included as a predictor (and zero otherwise), we used a k-means algorithm to cluster countries based on similarities and differences in aggregate predictors of conflict fatalities. The results of cluster analysis are shown in Figure 10, and in a map in Figure 11.

Eight clusters were detected which, going one step up the ladder of aggregation, can be further grouped in five main clusters. Among these five main clusters, we have two broad groups. One is made of the countries in clusters A, C and D, including overall 11 countries. For countries in these three clusters, predictors of conflicts are mainly structural factors and (in some cases) non-violent events – not past conflict, internal or external.

**[FIGURE 10 ABOUT HERE]**

The first group (cluster A in figure 11) is composed of Senegal and Sudan. Conflict in these countries is predicted almost entirely by VDEM-political factors and, in the case of Senegal, by GDELT non-violent events as well. Another small group (cluster C in Figure 11) includes Egypt, Mauritania and Rwanda. Even for countries from this group, conflict intensity is predicted by few classes of variables: WDI-structural factors alone predict conflict in Mauritania, supplemented by VDEM-political-institutional factors in the case of Rwanda and Egypt. For the latter, GDELT non-violent events also are significant predictors. Cluster D groups Chad, Congo, Cote d'Ivoire, Ethiopia, South Sudan, and Tunisia. Conflict in all these countries also is predicted by few classes of factors: WDI-structural factors and GDELT non-violent events. For Tunisia only, VDEM democracy-related factors (perhaps unsurprisingly) also predict conflict.

**[FIGURE 11 ABOUT HERE]**

Previous conflict or violent events are instead key predictors of conflict intensity for countries in the remaining clusters B and E. Cluster E includes Algeria, Angola, Burkina Faso, Central African Republic, Ghana, and Mozambique. Violent events – ACLED or GED – predict conflicts in all countries of this cluster. Conflict in all these countries is also predicted by WDI-structural factors and, with the exception of Burkina Faso, VDEM- institutional variables. In the case of Mozambique and Central Africa, GEDLT non-violent events are significant predictors too.

The largest cluster by far is cluster B, including Cameroon, Congo DR, Kenya, Libya, Mali, Niger, Nigeria, Somalia, Uganda. In these countries, conflicts are predicted by diverse factors. All countries share, among relevant predictors, GDELT non-violent events, WDI-structural factors, VDEM-political factors and (except for Somalia, Kenya and Congo DR), ICGCW violent events. For each country, the forecasting model also retained two or more additional significant predictors (see Figure 10). In sum, countries in this group share a core set of conflict predictors, but they also share the fact that conflict is predicted by a range of diverse additional classes of factors.

A proper analysis of these clusters is beyond the aim of this paper. What is relevant here is that – whatever the clustering approach taken – patterns shared by groups of countries do emerge when looking at similarities and differences in terms of conflict predictors, which permits to reconnect by-country models to more general tendencies.

## 5. Conclusions

This article illustrated the conceptual underpinnings, data, methods, and results of DynENet, an approach to forecasting conflict intensity that we first implemented in the context of the prediction competition organised by the ViEWS project (Hegre et al. 2022; Vesco et al. 2022). It also presented an evaluation of the performance of our models, as well as an illustration of how the results can be used to interpret and analyze the configuration of conflict dynamics in different contexts.

DynENet brings some novel contributions to the design of conflict forecasting approaches. First, rather than designing a single forecasting model, we modelled conflict in each individual country separately. We took the complexity of conflicts and their drivers seriously, and did not assume that a single model can predict (or describe) conflicts in any context. Personalised modelling is taking hold in the medical sciences, but is not yet common in the context of conflict modelling – and barely used in forecasts of social processes either (but see Carammia et al. 2022). Second, we employed a dynamic model that proved both suitable and computationally efficient enough to adapt to the complexity of conflict dynamics over space and time, allowing us to include (and select among) a large number of variables – about 750 indicators, including several data features provided by the ViEWS project as well as about 100 variables from the GDELT event dataset.

On top of the ViEWS dataset, we added GDELT event data to our models. GDELT did not markedly improve the reliability of forecasts in the test set; for some countries, they even made forecasts slightly less accurate. However, within the framework of our approach, the inclusion of event data carries some qualitative benefits, as it contributes to the third novel element of our approach – the interpretability of forecasts. Because our dynamic model selects the variables relevant to each country/month, and drops the others, the retained predictors and their varying importance can be analyzed to describe the varying drivers of conflict over space and time. The results also highlighted that some regularities exist even in single conflict processes, and that our model was able to capture them. The article presented four such forecast interpretations, and gave an indication of the qualitative insights that they can bring to understanding of the correlates of conflicts. We also showed that some general, cross-country tendencies emerge if countries are clustered based on their similarities and differences in terms conflict predictors – thus reconnecting country-specific processes to broader international regularities in conflict processes.

As discussed extensively in Vesco et al. (2022) our approach performed well in the test set compared to eleven other models – including nine models taking part to the competition, a benchmark Random Forest model, and to a no-change scenario. It ranked fourth in terms of MSE, second in terms of predicted variance compared to actual values, and first (or second, depending on the forecasting step) for contribution to an ensemble comprising all models. DynENet was relatively less efficient forecasting the true future. Still, it performed well above the median performance of competing models on twelve out of thirteen evaluation metrics; much closer to the best-performing model than the median in most metrics; and first or second in terms of (standard deviation or mean) forecast calibration.

However, this is a first implementation of a novel forecasting approach, and several avenues exist for possible improvements. An important limitation of our approach is that it was not able to generate forecasts for a significant number of countries (17 out of 52). This was due to either no past history of conflict or too sparse data in the dependent variable. Above, a solution to this issue has been proposed based on matching covariates of countries for which the estimation was possible to those for which the estimation failed, and then applying the estimated model to these latter countries.

GDELT event data gave a limited contribution to the effectiveness of our forecasts; in some cases, it even made them slightly less efficient. However, their contribution should be assessed on a case-by-case basis, and as part of a trade-off between a possible (marginal) gain or loss in forecasting performance on the one side, and a contribution to the interpretation of the processes underlying conflicts on the other. Still, future extensions of this or similar models should also consider testing alternative and more established sources of event data, such as the UTD event dataset (Kim et al., 2019; Solaimani et al., 2016).

Several technical improvements to the modelling approach also are possible. First, training and calibration windows could be more accurately calibrated. Second, the model could be made even more dynamic by training it on moving time windows (cf. Carammia et al. 2022), similar to what was already done here with the ex-post variable importance analysis. Such changes to model training and calibration are likely to further improve forecasting performance. Third, the computational efficiency of the model, and its very design based on variable selection, could be exploited to include more classes of predictors. In future developments we plan to add both traditional data on arms expenditures as well as non-traditional data on web searches. These could improve both forecasting accuracy and breadth of interpretability.

Finally, the architecture of the model and the use of event and other near-real time data would make it possible to add a preliminary layer of 'early warning' analysis to generate a number of statistics on stability, volatility and change points for single variables (cf. Carammia et al. 2022). The system could then be running in a 'live' version (or repeatedly actioned over time), and trigger alerts when detecting significant changes in the trends of individual series. Past correlations between single covariates and the outcome variable, as well as the lags that maximise such correlations, could be estimated.

The lagged predictors estimated in this way could be incorporated in the subsequent forecasting step, which we expect could further improve the predictive performance of the model.

## Acknowledgments

We are grateful to Francesco Bonaccorso for excellent research assistance, and to two anonymous reviewers for their comments that greatly improved the manuscript. We would also like to thank Mike Colaresi, Håvard Hegre, and Paola Vesco for organizing the Violence Early Warning System prediction competition and workshop, and all competition and workshop participants.

## References

Bell, Curtis. 2016. *The Rulers, Elections, and Irregular Governance Dataset (REIGN)*. Broomfield, CO: OEF Research. oefresearch. org.

Brandt, Patrick T., John R. Freeman, and Philip A. Schrodt. 2011. "Real Time, Time Series Forecasting of Inter- and Intra-State Political Conflict." *Conflict Management and Peace Science* 28 (1): 41–64. https://doi.org/10.1177/0738894210388125.

Buhaug, Halvard, Jack S Levy, and Henrik Urdal. 2014. "50 Years of Peace Research: An Introduction to the Journal of Peace Research Anniversary Special Issue." *Journal of Peace Research* 51 (2): 139–44. https://doi.org/10.1177/0022343314521649.

Carammia, Marcello, Stefano Maria Iacus, and Teddy Wilkin. 2022. "Forecasting Asylum-Related Migration Flows with Machine Learning and Data at Scale." *Nature Scientific Reports*, Social Physics, 12: 1–16. https://doi.org/10.1038/s41598-022-05241-8.

Colaresi, Michael, and Zuhaib Mahmood. 2017. "Do the Robot: Lessons from Machine Learning to Improve Conflict Forecasting." *Journal of Peace Research* 54 (2): 193–214. https://doi.org/10.1177/0022343316682065.

Coppedge, Michael, John Gerring, David Altman, Michael Bernhard, Steven Fish, Allen Hicken, Matthew Kroenig, et al. 2011. "Conceptualizing and Measuring Democracy: A New Approach." *Perspectives on Politics* 9 (2): 247–67. https://doi.org/10.1017/S1537592711000880.

Denny, Elaine K, and Barbara F Walter. 2014. "Ethnicity and Civil War." *Journal of Peace Research* 51 (2): 199–212. https://doi.org/10.1177/0022343313512853.

D'Orazio, Vito. 2021. "Conflict Forecasting and Prediction." In *Oxford Research Encyclopedia of International Studies*. Oxford: Oxford University Press. https://doi.org/10.1093/acrefore/9780190846626.013.514.

Gerner, Deborah J., Philip A. Schrodt, Rajaa Abu-Jabr, and Ömür Yilmaz. 2002. "Conflict and Mediation Event Observations (CAMEO): A New Event Data Framework for the Analysis of Foreign Policy

Interactions." Presented at the Annual Meeting of the International Studies Association, New Orleans.

Goldstein, Joshua S. 1992. "A Conflict-Cooperation Scale for WEIS Events Data." *Journal of Conflict Resolution* 36 (2): 369–85. https://doi.org/10.1177/0022002792036002007.

Hastie, Trevor, Robert Tibshirani, and Jerome Friedman. 2009. *The Elements of Statistical Learning: Data Mining, Inference, And Prediction*. 2nd ed. Springer Series in Statistics. Springer.

Hegre, Håvard. 2014. "Democracy and Armed Conflict." *Journal of Peace Research* 51 (2): 159–72. https://doi.org/10.1177/0022343313512852.

Hegre, Håvard, Marie Allansson, Matthias Basedau, Michael Colaresi, Mihai Croicu, Hanne Fjelde, Frederick Hoyles, et al. 2019. "ViEWS: A Political Violence Early-Warning System:" *Journal of Peace Research*, February. https://doi.org/10.1177/0022343319823860.

Hegre, Håvard, Nils W Metternich, Håvard Mokleiv Nygård, and Julian Wucherpfennig. 2017. "Introduction: Forecasting in Peace Research." *Journal of Peace Research* 54 (2): 113–24. https://doi.org/10.1177/0022343317691330.

Hegre, Håvard, Paola Vesco, and Michael Colaresi. 2022. "Lessons from an Escalation Prediction Competition." *International Interactions* 48 (Special Issue on an escalation prediction competition).

Hoerl, Arthur E., and Robert W. Kennard. 1970. "Ridge Regression: Biased Estimation for Nonorthogonal Problems." *Technometrics* 12 (1): 55–67. https://doi.org/10.1080/00401706.1970.10488634.

Iacus, Stefano M., Gary King, and Giuseppe Porro. 2012. "Causal Inference without Balance Checking: Coarsened Exact Matching." *Political Analysis* 20 (1): 1–24. https://doi.org/10.1093/pan/mpr013.

Iacus, Stefano M., Gary King, and Giuseppe Porro . 2019. "A Theory of Statistical Inference for Matching Methods in Causal Research." *Political Analysis* 27 (1): 46–68. https://doi.org/10.1017/pan.2018.29.

Kathman, Jacob D., and Reed M. Wood. 2011. "Managing Threat, Cost, and Incentive to Kill: The Short- and Long-Term Effects of Intervention in Mass Killings." *The Journal of Conflict Resolution* 55 (5): 735–60. https://doi.org/10.1177/0022002711408006.

Kim, HyoungAh, Vito D'Orazio, Patrick T. Brandt, Jared Looper, Sayeed Salam, Latifur Khan, and Michael Shoemate. 2019. "UTDEventData: An R Package to Access Political Event Data." *Journal of Open Source Software* 4 (36): 1322. https://doi.org/10.21105/joss.01322.

Koubi, Vally. 2019. "Climate Change and Conflict." *Annual Review of Political Science* 22 (1): 343–60. https://doi.org/10.1146/annurev-polisci-050317-070830.

Leetaru, Kalev, and Philip A. Schrodt. 2013. "GDELT: Global Data on Events, Location, and Tone." In *ISA Annual Convention*.

Nordås, Ragnhild, and Nils Petter Gleditsch. 2007. "Climate Change and Conflict." *Political Geography*, Climate Change and Conflict, 26 (6): 627–38. https://doi.org/10.1016/j.polgeo.2007.06.003.

Orsini, Amandine, Philippe Le Prestre, Peter M Haas, Malte Brosig, Philipp Pattberg, Oscar Widerberg, Laura Gomez-Mera, et al. 2020. "Forum: Complex Systems and International Governance." *International Studies Review* 22 (4): 1008–38. https://doi.org/10.1093/isr/viz005.

Pettersson, Therése, and Magnus Öberg. 2020. "Organized Violence, 1989–2019." *Journal of Peace Research* 57 (4): 597–613. https://doi.org/10.1177/0022343320934986.

Raleigh, Clionadh, Andrew Linke, Håvard Hegre, and Joakim Karlsen. 2010. "Introducing ACLED: An Armed Conflict Location and Event Dataset: Special Data Feature." *Journal of Peace Research* 47 (5): 651–60. https://doi.org/10.1177/0022343310378914.

Sanín, Francisco Gutiérrez, and Elisabeth Jean Wood. 2014. "Ideology in Civil War: Instrumental Adoption and Beyond." *Journal of Peace Research* 51 (2): 213–26. https://doi.org/10.1177/0022343313514073.

Schneider, Gerald, Nils Petter Gleditsch, and Sabine C. Carey. 2010. "Exploring the Past, Anticipating the Future: A Symposium." *International Studies Review* 12 (1): 1–7. https://doi.org/10.1111/j.1468-2486.2009.00909.x.

Schrodt, Philip A. 1991. "Prediction of Interstate Conflict Outcomes Using a Neural Network." *Social Science Computer Review* 9 (3): 359–80. https://doi.org/10.1177/089443939100900302.

Solaimani, M., S. Salam, A. M. Mustafa, L. Khan, P. T. Brandt, and B. Thuraisingham. 2016. "Near Real-Time Atrocity Event Coding." In *2016 IEEE Conference on Intelligence and Security Informatics (ISI)*, 139–44. https://doi.org/10.1109/ISI.2016.7745457.

Sundberg, Ralph, and Erik Melander. 2013. "Introducing the UCDP Georeferenced Event Dataset." *Journal of Peace Research* 50 (4): 523–32. https://doi.org/10.1177/0022343313484347.

Tibshirani, Robert. 1996. "Regression Shrinkage and Selection Via the Lasso." *Journal of the Royal Statistical Society: Series B (Methodological)* 58 (1): 267–88. https://doi.org/10.1111/j.2517-6161.1996.tb02080.x.

Toft, Monica Duffy. 2014. "Territory and War." *Journal of Peace Research* 51 (2): 185–98. https://doi.org/10.1177/0022343313515695.

Vesco, Paola, Håvard Hegre, Michael Colaresi, Remco Bastiaan Jansen, Adeline Lo, Gregor Reisch, and Nils B. Weidmann. 2022. "United They Stand: Findings from an Escalation Prediction Competition." *International Interactions*, 1–37. https://doi.org/10.1080/03050629.2022.2029856.

Ward, Michael D, Brian D Greenhill, and Kristin M Bakke. 2010. "The Perils of Policy by P-Value: Predicting Civil Conflicts." *Journal of Peace Research* 47 (4): 363–75. https://doi.org/10.1177/0022343309356491.

Weidmann, Nils B., and Michael D. Ward. 2010. "Predicting Conflict in Space and Time." *Journal of Conflict Resolution* 54 (6): 883–901. https://doi.org/10.1177/0022002710371669.

World Bank. "World Development Indicators." Washington, D.C: The World Bank. https://datacatalog.worldbank.org/dataset/world-development-indicators.

Zou, Hui, and Trevor Hastie. 2005. "Regularization and Variable Selection via the Elastic Net." *Journal of the Royal Statistical Society: Series B (Statistical Methodology)* 67 (2): 301–20. https://doi.org/10.1111/j.1467-9868.2005.00503.x.

**Tables**

| Evaluation Metric | s2 | s3 | s4 | s5 | s6 | s7 |
| --- | --- | --- | --- | --- | --- | --- |
| MSE | 0.977 | 0.978 | 0.978 | 0.967 | 0.976 | 0.967 |
| TADDA–A | 0.995 | 0.992 | 0.991 | 0.983 | 0.986 | 0.982 |
| TADDA–B | 0.995 | 0.997 | 0.988 | 0.981 | 0.989 | 0.988 |

Table 1. Efficiency ratios of DynENet over LASSO, steps 2-7.

| Evaluation Metric | s2 | s3 | s4 | s5 | s6 | s7 |
| --- | --- | --- | --- | --- | --- | --- |
| MSE | 0.006 | 0.021 | 0.035 | 0.021 | 0.023 | 0.022 |
| TADDA–A | 0.007 | 0.012 | 0.020 | 0.013 | 0.009 | 0.014 |
| TADDA–B | 0.008 | 0.009 | 0.013 | 0.011 | 0.009 | 0.016 |

Table 2. Performance loss from removing GDELT data from the model, steps 2-7

|  | DynENet full | DynENet no-GDELT | DynENet full *minus* no-GDELT | LASSO | DynENet/ LASSO | DynENet no-GDELT/ LASSO |
| --- | --- | --- | --- | --- | --- | --- |
| Country | MSE | MSE | Data ablation loss | MSE | Efficiency ratio | Efficiency ratio |
| AO | 0.438 | 0.438 | 0 | 0.438 | 1 | 1 |
| BF | 6.453 | 6.566 | -0.113 | 6.695 | 0.964 | 0.981 |
| BI | 0.751 | 0.742 | 0.008 | 0.765 | 0.981 | 0.971 |
| BJ | 0.047 | 0.047 | 0 | 0.047 | 1 | 1 |
| CD | 1.453 | 1.524 | -0.071 | 1.637 | 0.887 | 0.931 |
| CF | 1.389 | 1.376 | 0.013 | 1.396 | 0.995 | 0.986 |
| CG | 0.039 | 0.039 | 0 | 0.039 | 1 | 1 |
| CI | 0.209 | 0.209 | 0 | 0.209 | 1 | 1 |
| CM | 1.233 | 1.155 | 0.077 | 1.620 | 0.761 | 0.713 |
| DJ | 0.000 | 0.000 | 0 | 0.000 | 1 | 1 |
| DZ | 3.318 | 3.296 | 0.022 | 3.190 | 1.040 | 1.033 |
| EG | 2.117 | 2.082 | 0.035 | 3.107 | 0.681 | 0.670 |
| ER | 0.000 | 0.000 | 0 | 0.000 | 1 | 1 |
| ET | 1.395 | 1.395 | 0 | 1.395 | 1 | 1 |
| GH | 0.013 | 0.013 | 0 | 0.013 | 1 | 1 |
| GM | 0.134 | 0.134 | 0 | 0.134 | 1 | 1 |
| KE | 1.705 | 1.638 | 0.067 | 1.750 | 0.975 | 0.936 |
| LS | 0.013 | 0.013 | 0 | 0.013 | 1 | 1 |
| LY | 4.856 | 4.613 | 0.243 | 4.785 | 1.015 | 0.964 |
| ML | 2.163 | 1.935 | 0.228 | 2.118 | 1.021 | 0.913 |
| MR | 0.002 | 0.002 | 0 | 0.002 | 1 | 1 |
| MW | 0.053 | 0.053 | 0 | 0.053 | 1 | 1 |
| MZ | 5.878 | 5.637 | 0.242 | 5.885 | 0.999 | 0.958 |
| NA | 0.034 | 0.034 | 0 | 0.034 | 1 | 1 |
| NE | 3.842 | 3.984 | -0.142 | 4.003 | 0.960 | 0.995 |
| NG | 1.332 | 1.325 | 0.007 | 1.268 | 1.050 | 1.045 |
| RW | 0.483 | 0.483 | 0 | 0.483 | 1 | 1 |
| SD | 4.570 | 4.543 | 0.027 | 4.698 | 0.973 | 0.967 |
| SN | 0.067 | 0.067 | 0 | 0.067 | 1 | 1 |
| SO | 0.461 | 0.454 | 0.007 | 0.459 | 1.004 | 0.989 |
| SS | 2.698 | 2.698 | 0 | 2.698 | 1 | 1 |
| TD | 4.398 | 4.276 | 0.123 | 4.348 | 1.011 | 0.983 |
| TN | 0.493 | 0.493 | 0 | 0.493 | 1 | 1 |
| UG | 0.341 | 0.343 | -0.002 | 0.349 | 0.977 | 0.982 |
| ZA | 0.118 | 0.118 | 0 | 0.118 | 1 | 1 |

Table 3. Country-level MSE at step 7 for different models, and resulting efficiency ratios and data ablation losses.

|  | *Forecasted trend* | *Forecasting performance in test set* | | | *Number and class of predictors* | |
|---|---|---|---|---|---|---|
| *Country* | *Cumulative delta* | MSE | DAL | DynENet/LASSO Efficiency ratio | *Number retained variables* | *Class retained variables* |
| Cameroon | 1.931 | 1.291 | 0.211 | 0.867 | 37 | GED (5) |
|  |  |  |  |  |  | WDI (16) |
|  |  |  |  |  |  | GDELT (10) |
|  |  |  |  |  |  | ACLED (4) |
|  |  |  |  |  |  | V-Dem (1) |
|  |  |  |  |  |  | FVP (1) |
| Nigeria | -3.907 | 0.915 | – | 0.816 | 48 | WDI (31) |
|  |  |  |  |  |  | V-Dem (12) |
|  |  |  |  |  |  | GED (2) |
|  |  |  |  |  |  | ACLED (2) |
|  |  |  |  |  |  | FVP (1) |
| Rwanda | 0.357 | 0.429 | 0 | 1 | 3 | GDELT (1) |
|  |  |  |  |  |  | GED (1) |
|  |  |  |  |  |  | V-Dem (1) |
| Senegal | 1.090 | 0.051 | 0 | 1 | 6 | WDI (4) |
|  |  |  |  |  |  | GDELT (2) |

Table 4. Summary statistics and predictors for Cameroon, Nigeria, Rwanda, Senegal.

# Figures

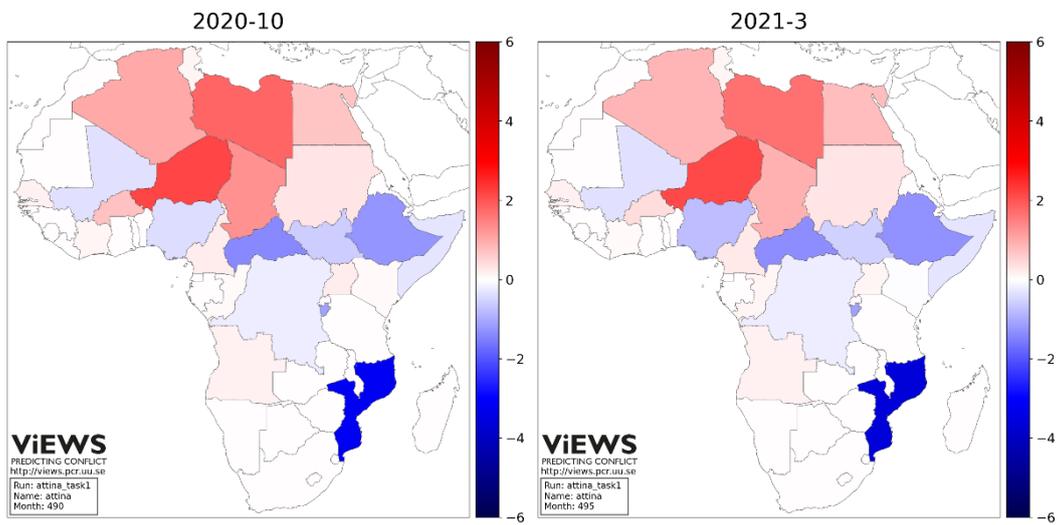

Figure 1. Predicted change at step 2 (October 2020) and step 7 (March 2021).

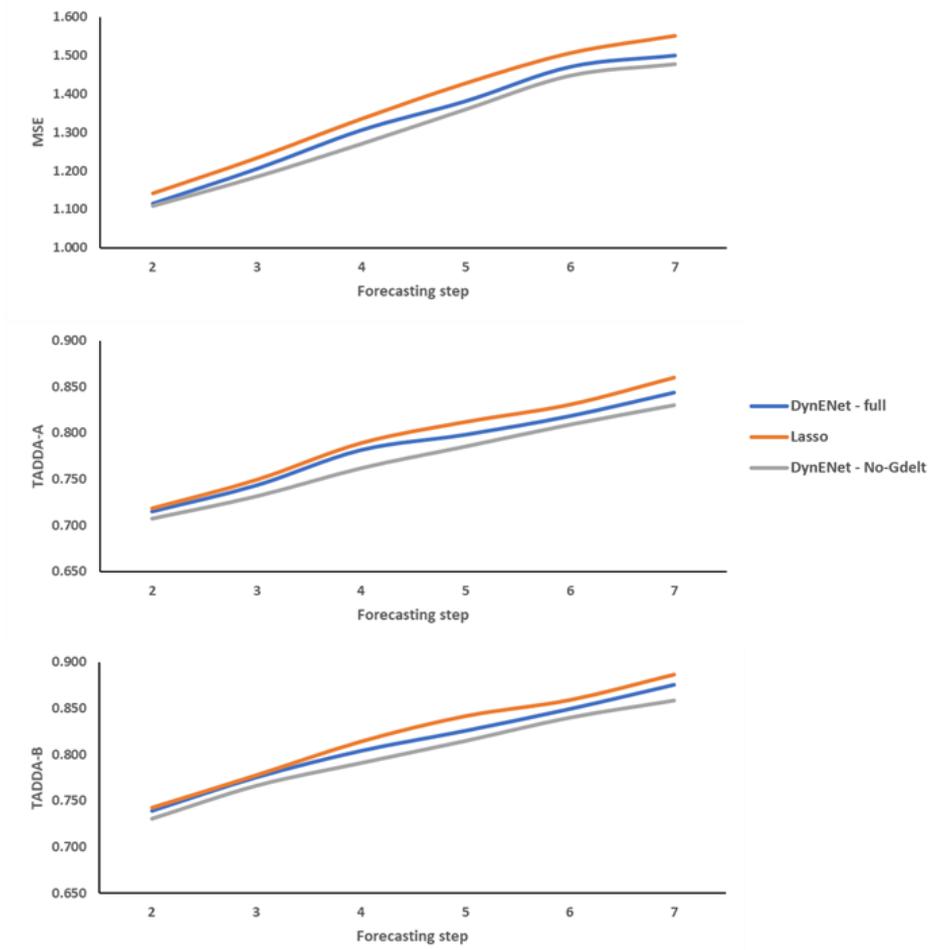

Figure 2. DynENet – full model, LASSO, DynENet – without GDELT data. Average forecasting error in test set (January 2017-December 2019), steps 2-7. MSE, TADDA-A, TADDA-B.

(a) 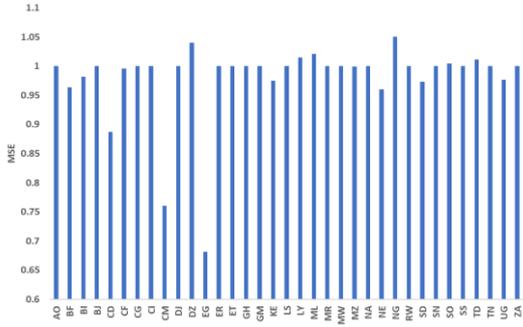 (b) 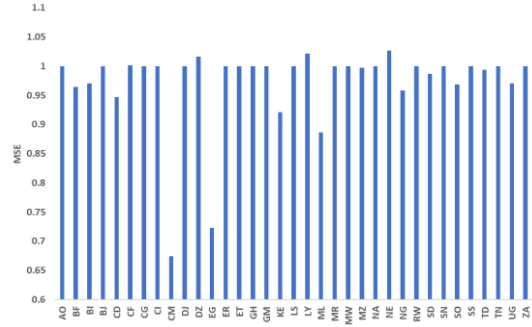

Figure 3. Average efficiency ratio of DynENet [full model (a) and without GDELT data (b)] over LASSO at step 7, for single countries.

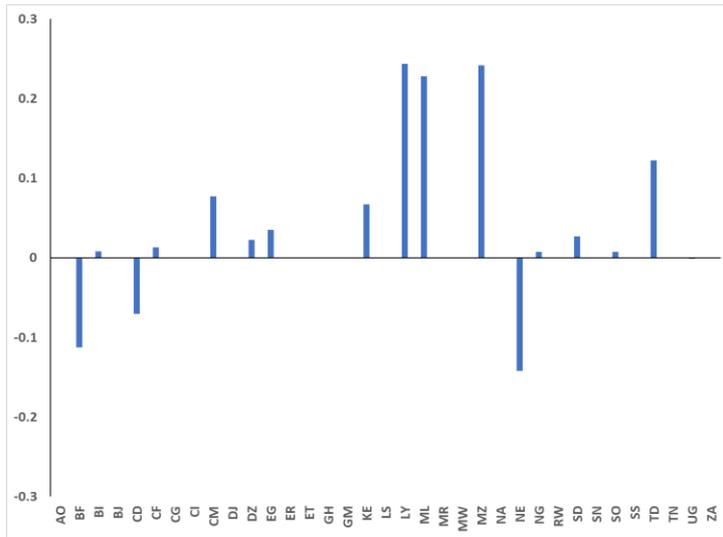

Figure 4. GDELT Data Ablation Loss at step 7, for single countries.

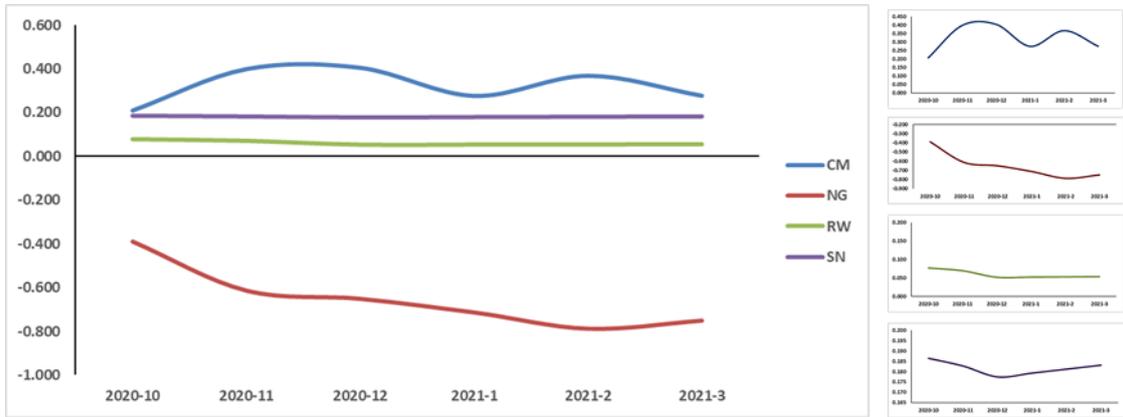

Figure 5. Forecasted monthly trend in conflict fatalities, steps 2-7 (October 2020 - March 2021). Single scale (left quadrant), and separate scales (right quadrants).

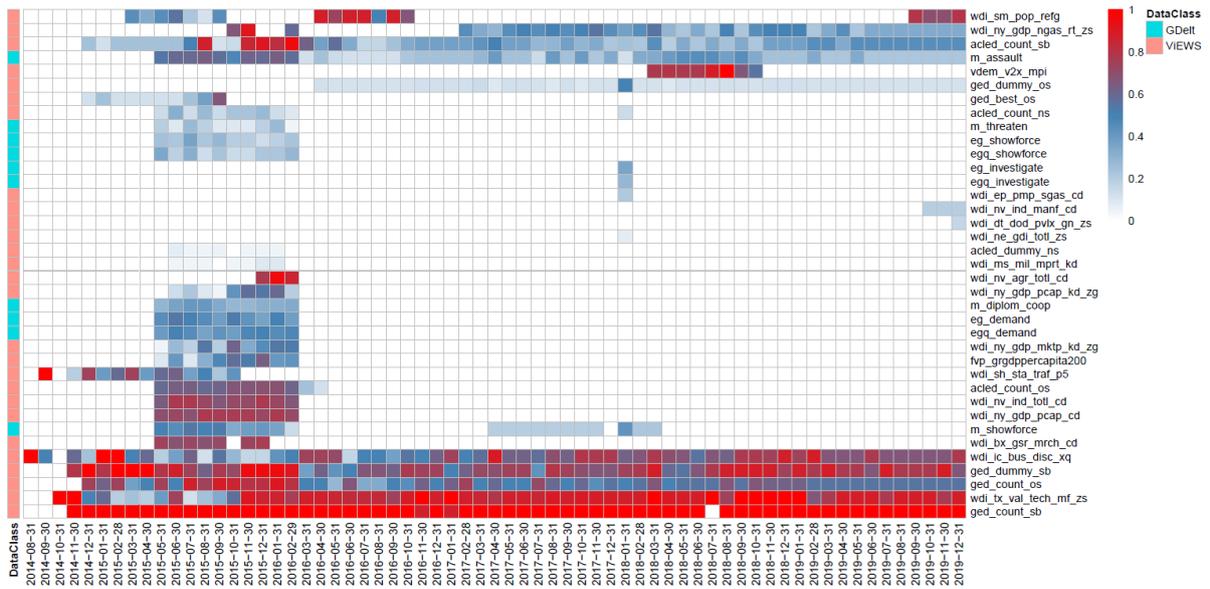

Figure 6. Predictors selected by DynENet and relative importance, Cameroon.

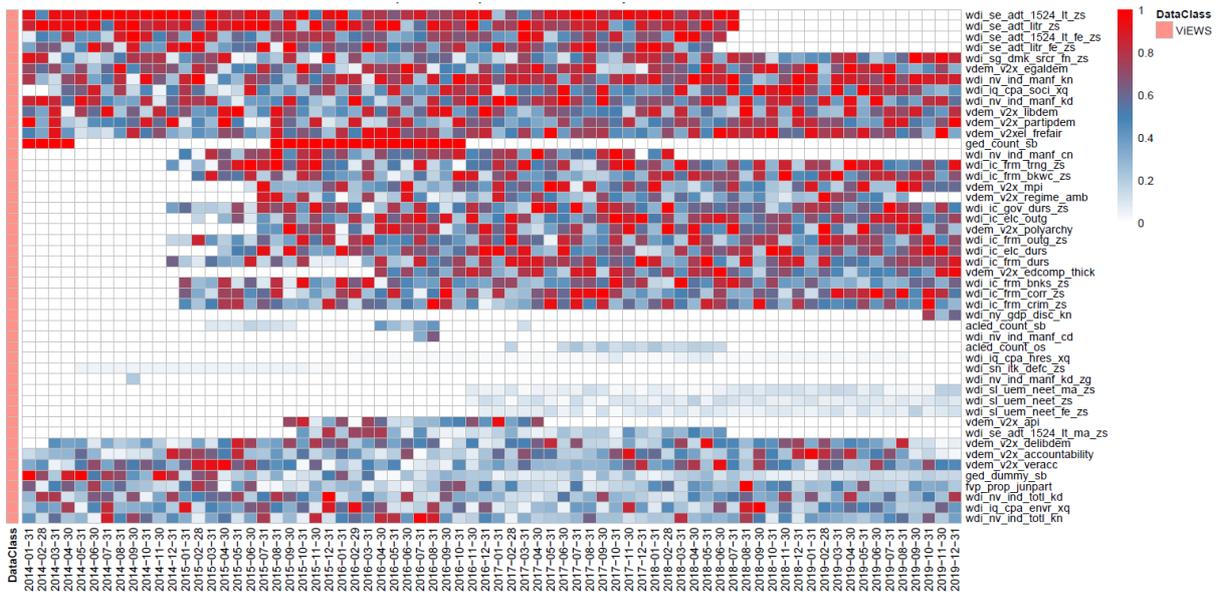

Figure 7. Predictors selected by DynENet and relative importance, Nigeria.

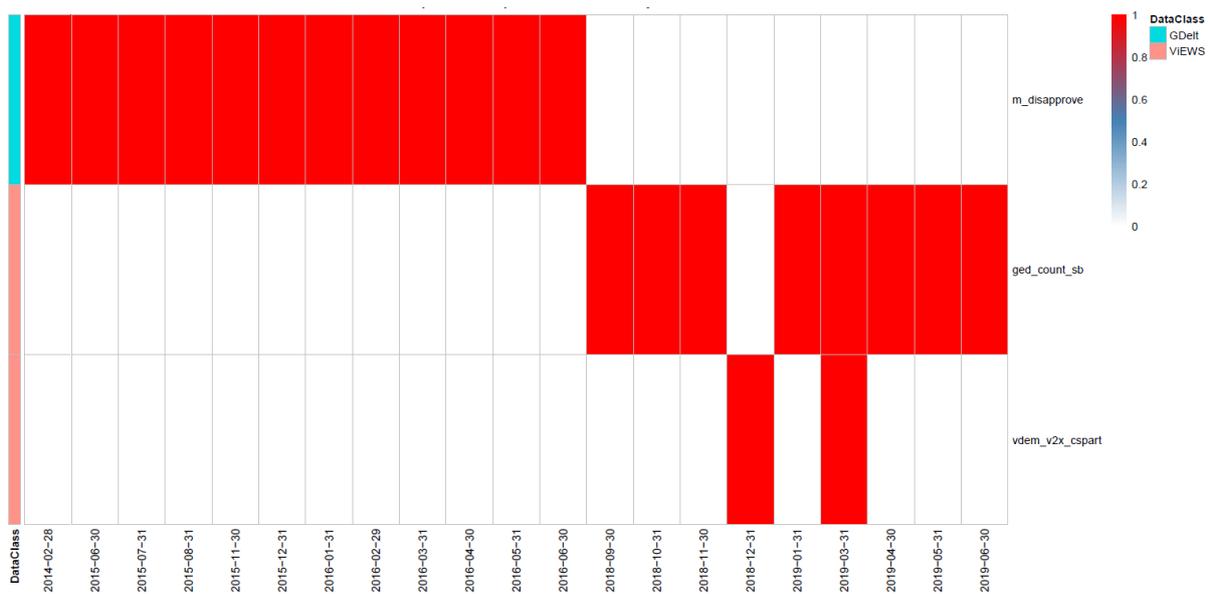

Figure 8. Predictors selected by DynENet and relative importance, Rwanda.

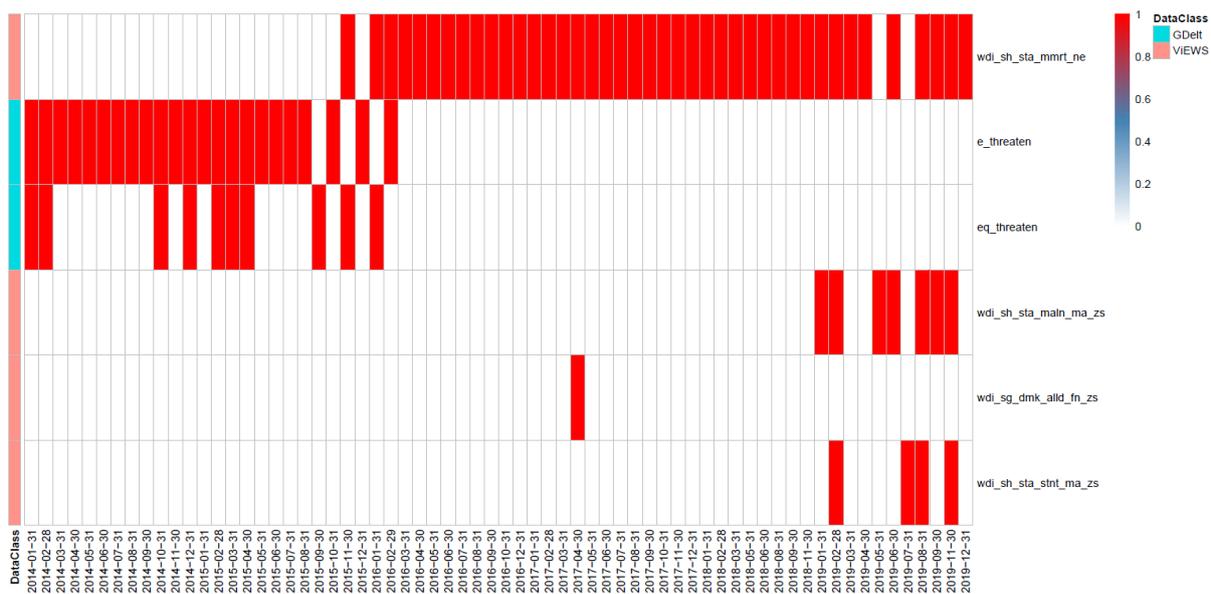

Figure 9. Predictors selected by DynENet and relative importance, Senegal.

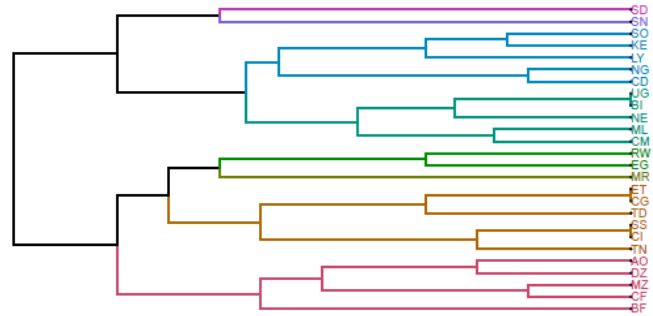
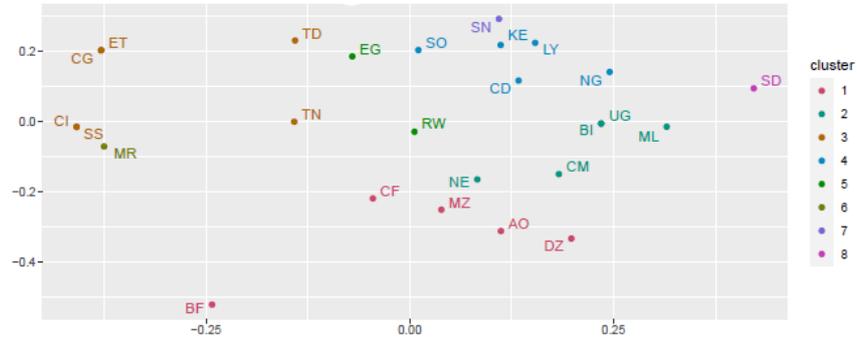
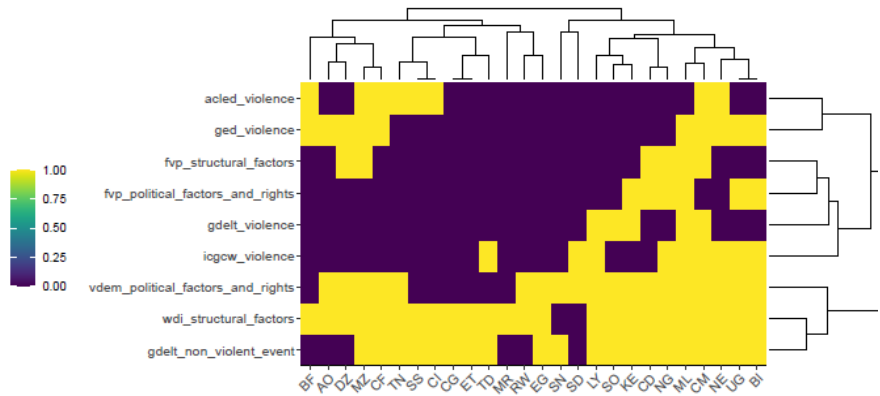

Figure 10. Clustering of countries by class (source and content) of conflict predictors.

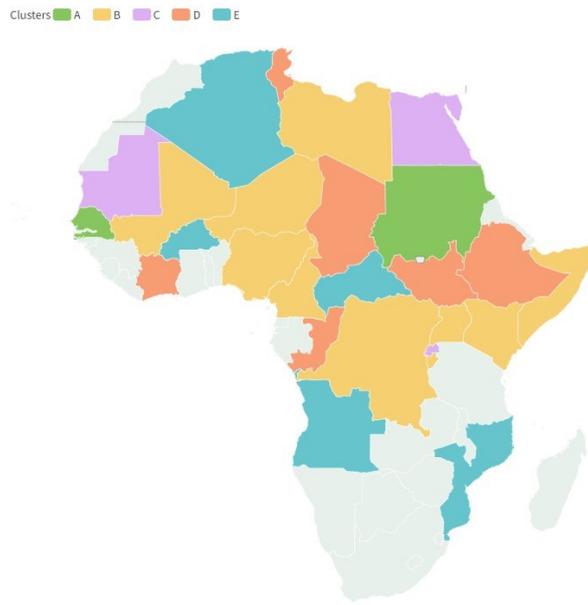

Figure 11. Map of country clusters